\documentstyle[aps,prl]{revtex}
\input{epsf}
\begin{document}
\title {Diffusive Boundary Layers in the Free-Surface Excitable Medium Spiral}
\author{David A. Kessler}
\address{Minerva Center and Dept. of Physics\\
Bar-Ilan University\\
Ramat Gan, Israel}
\author{Herbert Levine}
\address{Institute for Nonlinear Science \\
University of California, San Diego \\ La Jolla, CA  92093-0402}
\maketitle
\begin{abstract}
Spiral waves are a ubiquitous feature of the nonequilibrium dynamics of
a great variety of excitable systems. In the limit of a large separation
in timescale between fast excitation and slow recovery, one can reduce the
spiral problem to one involving the motion of a free surface separating the
excited and quiescent phases. In this work, we study the free surface
problem in the limit of small diffusivity for the slow field variable.
Specifically, we show that a previously found spiral solution in
the diffusionless limit can be extended to finite diffusivity, without
significant alteration. This extension involves the creation of a 
variety of boundary layers which cure all the undesirable singularities
of the aforementioned solution. The implications of our results for
the study of spiral stability are briefly discussed.
\end{abstract}

Understanding the behavior of spiral waves in excitable media remains an 
important and
challenging problem~\cite{ref1}.  From the perspective of numerical simulation, two 
component reaction-diffusion models~\cite{ref2,ref3} have been shown to capture the important features of 
spiral patterns, in
particular the transition from rigid rotation to meandering and the 
phenomenology of the
nonlinear meandering state~\cite{ref4,ref5}.  More recently,
simulations~\cite{ref6},  and a numerical stability analysis~\cite{ref7} 
based on the 
free boundary limit~\cite{ref7'} (see
below) have confirmed that the finite thickness of the front (separating 
the excited region from the
quiescent one) is not crucial for the meandering behavior.  It is 
therefore of interest to pursue
analytical techniques (which make use of this free boundary limit) in the 
hope of gaining a more
fundamental understanding of the nature of this instability.

This work reports progress towards the aforementioned goal of having an 
analytic theory of spiral
waves.  Specifically, we revisit an approach due to Pelce and Sun~\cite{ref8} who 
derived a spiral solution in
the case of zero diffusion for the ``controller" variable.  Their 
solution exhibits singular behavior
near the spiral ``tip" (for example, the front curvature has a 
discontinuous derivative across the tip
point) which  raises questions regarding the validity of 
the solution and to date has precluded
a full stability analysis.  Here, we show how the inclusion of small but 
finite diffusivity provides,
via the introduction of boundary layers, for a regularization of the 
singular behaviors.  This
therefore confirms the physical validity of their construction.  The 
implications of our results for a
(future) calculation of spiral stability are discussed at the end.

The free boundary approach starts from the equations coupling a 
concentration field $v
(\vec{x}, t)$ to an interface between an excited state of the medium 
(``+") and the quiescent
state (``-").  In so-called ``Fife''-scaled units~\cite{ref10},
\begin{equation}
\frac{\partial v_\pm}{\partial t} = g_\pm - \mu v_\pm + D \nabla^2 v
\end{equation}
where $\mu$ is a positive constant, $g_+$, $g_-$ are positive, negative 
constants and $v_\pm$ refers to 
the field in the +,-
regions respectively.  The field obeys the boundary conditions at the 
interface~\cite{ref9}
\begin{equation}
c_n + \kappa = - v_{int}
\end{equation}
where $c_n$ is the normal velocity, $\kappa$ the curvature, and the value 
of $v$ (as well as its normal derivative) is continuous
across the interface.  

This system of equations is not rigorously derivable from the original 
reaction diffusion system
\begin{equation}
\frac{\partial u}{\partial t} = \nabla^2 u + \frac{f(u, v)}{\epsilon}
\end {equation}
\begin{equation}
\frac{\partial v}{\partial t} =  D \nabla^2 v +  g(u, v)
\end{equation}
The reason for this is that replacing the ``propagator" field equation 
for $u(x,t)$ by an interfacial
boundary condition is only valid asymptotically as $\epsilon \rightarrow 
0$.  The coefficient $\mu$ of the linear term in eq. (1) is formally small, of 
order $\epsilon^{1/3}$. Hence we cannot rigorously keep this linear term 
without keeping additional terms of the same order as well.  
As we have already 
mentioned, however, simulations show that eqns. (1) and (2) 
do capture the spiral 
phenomenology of interest, at least for finite diffusivity $D$.  Furthermore,
an exact numerical steady-state 
solution~\cite{ref11} and subsequent stability analysis~\cite{ref7} directly 
supports this conclusion. On the other hand, results obtained by dropping
the linear term and working at small $D$ seem at present to be rather 
unphysical~\cite{ref10',ref7}.   We 
will therefore adopt eqns.
(1, 2) as our fundamental model and proceed to consider the small $D$ limit.

\begin{figure}
\centerline{\epsfxsize = 3.in \epsffile{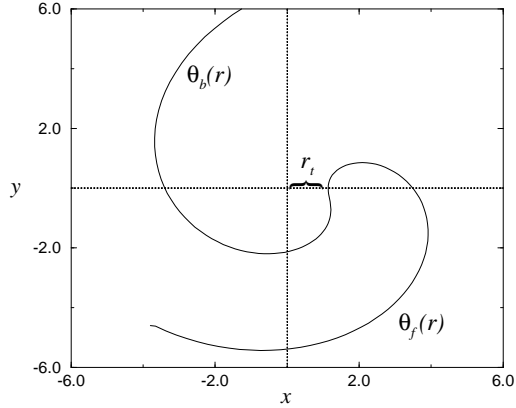}}
\caption{A typical spiral, with $r_t$, $\theta_f(r)$, $\theta_b(r)$ indicated}
\end{figure}

This free boundary problem with $D = 0$ was first tackled by Pelce and 
Sun~\cite{ref8}.  If one assumes a
uniformly rotating field $v(\vec{x}, t) \rightarrow v(r, \theta - \omega 
t)$ one obtains 
\begin{equation}
- \omega \frac{\partial v_\pm}{\partial \theta} = g_\pm - \mu v
\end{equation}
This can be solved by imposing the matching conditions $v_+ (\theta_f) = 
v_-(\theta_f)$ and
$v_+ (\theta_b) = v_-(2\pi + \theta_b)$ where $\theta_f(r)$ and $\theta_b(r)$ 
are the positions of the
interface from quiescent to excited and back to quiescent, as a function of the
radius $r$ (see Fig. 1).  Defining $\tilde \omega \equiv \omega/\mu$, 
${\tilde g}_\pm \equiv g_\pm/\mu$, the
solution can be written as 
\begin{equation}
v_\pm (r, \theta) = {\tilde g}_\pm + A_\pm (r) e^{\frac {\theta}{\tilde\omega}}
\end{equation}
where
$$
A_-(r) = ({\tilde g}_+ - {\tilde g}_-) \frac{(e ^{-\theta_f / {\tilde\omega}} - 
e^{- \theta_b / \tilde \omega})}{1 -
e^{2\pi / {\tilde \omega}}}$$ 
$$
A_+ (r) = A_- (r) - ({\tilde g}_+ - {\tilde g}_-) e^{- \theta_f / {\tilde \omega}}
$$
This can then be substituted into the ``eikonal" equation (2) to find the 
interfaces $\theta_f (r),
\theta_b (r)$.  Near the tip $(\theta = 0, r= r_{t})$, smoothness 
requires $- \theta_b (r) =
\theta_f (r) = \alpha \sqrt{r - r_{t}}$, for some positive $\alpha$.  
This condition allows for the 
determination of the
rotation frequency $\omega$.  It is easy to check that the field value 
$v$ takes the value $\tilde{g}_-$ at the tip and for all $r < r_{t}$.

This solution has several unattractive features.  Since the interface 
positions $\theta_{f,b}$ vary as
$\sqrt{r - r_{t}}$,
\begin{equation}
\frac{\partial v}{\partial r} \sim \frac{1}{\sqrt{r - r_{t}}}
\end{equation}
\noindent
for all $\theta$ when $r$ is close to but above the tip radius $r_{t}$.  Also, 
$\frac{\partial v}{\partial 
\theta}$ has a finite jump
discontinuity at $r=r_{t}$, $\theta=0$, so that the derivative of $v$ 
along the interface 
has a jump discontinuity at the tip where the front and back meet;
this leads via the
relation (2) to a similar jump for the derivative of the curvature.  In
addition, the normal derivative of $v$ has a finite jump discontinuity
across the interface.
These difficulties 
have had the practical effect of
making it impossible to do a full stability analysis of the spiral 
solution~\cite{ref12} and in general raises questions concerning
whether the introduction of finite diffusivity dramatically 
alters the conclusions one
derives via this construction.   We now show that in fact finite 
diffusivity smoothes out this
singularities without making any quantitatively significant change in the 
interface shape and
selected rotation frequency.

To proceed, we use an integral equation formulation of the 
problem~\cite{ref7'}.  The field 
$v - {\tilde g}_\pm$ obeys a homogeneous, linear field equation with 
a fixed discontinuity
across
the interface.  Following standard manipulations, we can therefore express 
$v$ in the rotating frame as
\begin{equation}
v_\pm (r,\theta) = {\tilde g}_\pm + ({\tilde g}_+ - {\tilde g}_-)
D \int ds^\prime \ \left( \hat{n}^\prime  
\cdot \vec{\nabla}^\prime
G - \frac{\omega r^\prime}{D} \hat{n}^\prime_\theta G \right)
\end{equation}
where $G$ is the Green's function for the equation
\begin{equation}
(D\vec{\nabla}^2 + \omega \frac{\partial}{\partial \theta} - \mu)\  G (r, 
\theta; r^\prime,
\theta^\prime)
= \frac{-\delta (r - r^\prime) \delta (\theta - \theta^\prime)}{r^\prime}
\end{equation}
An explicit representation of $G$ which will be used below is 
\begin{equation}
G(r, \theta; r^\prime, \theta^\prime) = \int^t_{-\infty} 
\frac{dt^\prime}{4\pi D (t-t^\prime)} \exp \{
\frac{-(r^2 + {r '} ^2 - 2rr^\prime \cos{ (\theta - \theta^\prime + 
{\omega} (t - t^\prime))} )}{4D (t -
t^\prime)}\  - \mu (t - t^\prime) \}
\end{equation}
The integral in (8) is over the entire interface with arclength variable 
$s^\prime$ and $\hat{n}^\prime$
is
the unit normal which points outward from the excited region; 
$\hat{n}^\prime_\theta$ is the
component of $\hat{n}$ in the direction of the $\hat{\theta}^\prime$ 
vector.  For a given
interface,
this construction gives a field which obeys the field equation and the 
continuity condition; a full
solution can then be found in principle by substituting the field value 
at the interface into the
eikonal condition and iterating the interface shape until the equation is 
satisfied.  What we will do
instead is evaluate this integral for the Pelce-Sun (PS) interface obtained at $ D = 
0$.  We will show
that
this field is completely free from singularities but differs from the 
Pelce-Sun field solution only by
terms which are small in the small $D$ limit.

Let us first focus on some point away from either the interface or $r = 
r_{t}$.  In this region, we will show how our formulation reproduces
the known results of Pelce and Sun. Since $G$ will
never be singular, the $\hat{n}^\prime \cdot \nabla^\prime G$ in (6) is 
order $D$ and can be
neglected.  The integral $\hat{n}^\prime_\theta ds^\prime$ can be 
replaced by $\pm dr^\prime$
(for
front and back respectively).  Now, the integrals over $r^\prime$ (in eq. 
6) and $t^\prime$ (in the
definition of $G$) are dominated by the saddle point contributions coming 
from
$$
r^\prime = r + 0 (\sqrt{D}) $$ $$
t^\prime_n = t + (\theta - \theta^\prime - 2\pi n) / {\omega} 
+ 0 (\sqrt {D})
$$
If $\theta > \theta^\prime$, $n$ runs from 1 to $\infty$; otherwise $n$ 
goes from 0 to $\infty$. 
Doing the Gaussian integrals around these saddle points leads after some 
algebra to
\begin{equation}
v_\pm = {\tilde g}_\pm - ({\tilde g}_+ - {\tilde g}_-) \left( \sum_n \exp (\theta - 
\theta_f - 2\pi n)/ {\tilde\omega}  -
\sum_n
\exp (\theta - \theta_b - 2\pi n)/ {\tilde\omega} \right)
\end{equation}
where the $n$ values in each sum obey the aforementioned rule.  
In the excited
region, for example, $\theta_f > \theta > \theta_b$, yielding
\begin{equation}
v_+ (\theta) = {\tilde g}_+ - ({\tilde g}_+ - {\tilde g}_-) e^{ \frac{\theta}{\tilde\omega} } \left[ 
e^{-\theta_f/ {\tilde\omega}}
\sum_{n=0}^\infty e^{-2\pi n/ {\tilde\omega}} - e ^{-\theta_b /{\tilde\omega}} 
\sum_{n=1}^\infty e ^{-2\pi n / {\tilde\omega}} \right]
\end{equation}
with a similar expression obtainable for $v_- (\theta)$.  Doing the sums, 
we directly recover the result of eq. (6) above.

We now wish to understand the departure from the Pelce-Sun solution due
to finite diffusivity. As we have seen, there are a number of
different regions where the Pelce-Sun solution breaks downs and exhibits
singular behavior. These breakdowns are all cured by boundary-layers
whose width vanishes with $D$. Interestingly enough, as we shall see, the
scaling of the boundary-layer width differs in each region. We first
examine the region $r \simeq r_{t}$, but 
still away from the
interface, i.e. $\theta$ is not close to 0.  
We saw in eq. (8) that $dv/dr$ was singular in this region.
Examing our integral representation for $v$ was see that the saddle point 
contribution from the $t^\prime$ integration 
is unchanged; however,
the Gaussian integral with $r^\prime \simeq r + 0( \sqrt{D})$ 
cannot be blithely extended to
infinity since the lower limit of the $r^\prime$ integration range is 
$r_{t}$.  Also, we must use
the fact that the integral of the Green's function along the front and 
back have opposite signs
and cancel to lowest order for $r^\prime \simeq r_{t} $.  Those 
considerations lead to the
expression
\begin{equation}
v_- (\theta) = {\tilde g}_- -  ({\tilde g}_+ - {\tilde g}_-) \omega r_{t}
{\sqrt{D}} \int^\infty_0 d \tilde{r}^\prime \ \frac{\partial G}{\partial 
\theta^\prime} (\tilde{r},
\tilde{r}^\prime ; \theta, \theta^\prime) \mid_{\theta^\prime = 0} 
(\theta_f (\tilde{r}^\prime) - \theta_b
(\tilde{r}^\prime) )
\end{equation}
where
\begin{equation}
 r = r_{t} + \sqrt{D} \tilde{r} ,\;\;\; r^\prime = r_{t}
 + \sqrt{D} 
\tilde{r}^\prime .\nonumber \end{equation}
Plugging in the expression of the Green's function and performing the $t'$ integral,
we obtain
$$
v_- (\theta) =  {\tilde g}_- + 2({\tilde g}_+ - {\tilde g}_-) \alpha D^{3/4} 
\frac{\partial F} {\partial \theta}$$
with 
\begin{equation}
F(\theta ) \ =  \  \sum_n \int_0^\infty \frac{d 
\tilde{r}^\prime \ \sqrt{\tilde{r}^\prime}}{\sqrt{4\pi D (2\pi n 
- \theta) / \omega}}  \exp \left[  \frac 
{-({\tilde{r}}^2 + {{\tilde r}{}'}^{2} -
2\tilde{r}
\tilde{r}')}{4 (2\pi n - \theta) / {\omega}} + ( \theta - 2\pi n)/ {\tilde\omega} 
\right] 
\end{equation}
$F$ can be expressed in terms of the parabolic cylinder 
function~\cite{ref13}
{\bf D}$_{\nu}$ for index $\nu = -3/2$:
\begin{eqnarray}
v_- (\theta) = & {\tilde g}_- + \frac { {\tilde g}_+ - {\tilde g}_-}{\sqrt{2}}
 D^{1/4} 
\frac{\partial}{\partial \theta} \left(
\sum_n \frac{\delta_n}{\mu}^{1/4} \exp ( \mu \delta_n/2) \right. \nonumber \\
& \ \ \ \ \cdot \left.
\exp \left( \frac {-
\tilde{r}^2}{4\delta_n} \right)  \mbox{{\bf D}}_{-3/2} \left( - 
\tilde{r} / \sqrt{\delta_n} \ \right) \right)
\end{eqnarray}
with $\delta _n \ = \ 2 (2 \pi n - \theta) / {\omega}$.
This structure represents a boundary layer of (minimum)
width $\sqrt {{2D |\theta| \over \omega}}$.  For negative $\tilde{r}$, the 
parabolic cylinder function decays as 
a Gaussian and $v_-(0)$ approaches ${\tilde g}_-$; 
for large positive $\tilde{r}$, ${\bf D}$ grows and cancels 
the exponential factor,
leading to the expected behavior $v_- (\theta) - {\tilde g}_- \sim \sqrt 
{\tilde{r}} D^{1/4} \sim \sqrt{r- r_{t}}$. 
Note that the change in the field itself
is negligible as it vanishes in the small $D$ limit as $D^{1/4}$, even though
the derivative is of order $D^{-1/4}$ and approaches infinity.

So, we have shown how integrating over the Pelce-Sun interface with the 
finite $D$ Green's
function leads to a regularization of the infinite slope discontinuity at 
$r =  r_{t}$, without
modifying the field. In fact, this lack of field modification is true
everywhere including the interface. This follows from the fact that we have
already shown that our
field construction agrees with the PS field everywhere that is a
distance O(1) away from the 
interface and the PS fields are continuous across the interface.
Field derivatives near the interfaces,
on the other hand, are significantly modified due to the presence of
diffusion-induced interfacial boundary layers.
We now turn to a discussion of the form of
these layers.  

Let us consider
first the interface ${\em away}$ from the tip.  Expressed in terms of our 
discussion so far, the only
necessary modification is to the $n = 0$ saddle point in the $t^\prime$ 
integral.
Specifically, if $\theta - \theta_f (r) = \tilde{\theta}$ is small 
(i.e. we are close to the front), we need to keep an extra term in the argument of the cosine 
appearing in the Green's function
\begin{equation}
\label{thetaeq}
\theta - \theta_f (r^\prime) + \omega (t - t^\prime)\  \simeq \
\tilde{\theta} + \frac{\partial
\theta_f}{\partial r^\prime} ( r - r^\prime) + \omega (t - t^\prime)
\end{equation}
The modified $t'$ saddle point occurs at
\begin{equation}
\omega (t - t^\prime) = - \tilde{\theta} - \frac{\partial 
\theta_f}{\partial r^\prime} (r - r^\prime)
\end{equation}
(which must of course be positive).  Doing the $t^\prime$ integral leaves 
us with an expression of the form
\begin{equation}
\int \frac{d r^\prime} {\sqrt {\frac{4 \pi D}{\omega} \mid \tilde{\theta} 
+ \frac{\partial
\theta_f}{\partial r} (r - r^\prime) \mid }} \exp \left [ \frac{- 
(r - r^\prime)^2}{({4D \over \omega} \mid
\frac{\partial \theta_f}{\partial r} ( r - r^\prime) + \tilde{\theta} 
\mid )} +( \tilde{\theta} +
\frac{\partial \theta_f}{\partial r^\prime} ( r - r^\prime)) / {\tilde \omega} \right]
\end{equation}
where the integral ranges over those values of $r'$ consistent with
the aforementioned positivity constraint on $t-t'$. This expression implies there is a crossover 
in behavior from the previous off-interface 
structure when $\tilde{\theta}$ is
order $D$ and that the range of relevant $r'$ variation is also of
order $D$.  The overall structure of the $v$ field can be 
shown to be~\cite{ref14}
\begin{equation}
v = v_{ps} + D \hat{v} \left( \frac{\tilde{\theta}}{D} \right)
\end{equation} and $\hat{v}$ is constant on one side, exponentially
decaying on the other side, of the interface.  
This form is necessary to cure the finite slope discontinuity across the PS 
interface.

In fact, the width of this interfacial 
boundary layer is proportional to $|\frac{\partial \theta_f}{\partial r}|$.
Since this derivative diverges near the tip, the boundary layer is
much wider than order $D$ in this region. To analyze this region, we modify
Eq. (\ref{thetaeq}) above, making explicit the square root dependence of
$\theta_f$ with $r^\prime$ near the tip:
\begin{equation}
\theta - \theta_f (r^\prime) + \omega (t - t^\prime) \simeq 
\tilde{\theta}- \alpha {\sqrt {r^\prime -
r_{t}}} + \omega (t - t^\prime)
\end{equation}

\noindent
(where $\tilde{\theta}$ is measured from zero angle).  Letting $r = 
r_{t} + \tilde{r},\  r^\prime
= r_{t} + \tilde{r}^\prime$, we obtain for  
the $G$ term
\begin{equation}
\frac{- \partial}{\partial \theta} \int \frac{d \tilde{r}^\prime \   2 \alpha
\sqrt{\tilde{r}^\prime}}  {\sqrt{4 \pi D ( - \tilde{\theta} + \alpha 
\sqrt{\tilde{r}^\prime })}} \exp
\left[ \frac{- (\tilde{r} - \tilde{r}^\prime)^2} {4D (- \tilde{\theta} + 
\alpha
\sqrt{\tilde{r}^\prime) } } + (\tilde{\theta} - \alpha 
\sqrt{\tilde{r}^\prime}) / {\tilde\omega} \right]
\end{equation}
The crossover from the previous $r \sim r_{t}$ behavior at finite 
$\tilde{\theta}$ occurs at 
\begin{equation}
\tilde{\theta} \sim \sqrt{\tilde r} \sim D^{1/3}
\end{equation}
\noindent
so that the boundary layer size in $\tilde r$ is
$\tilde{r}^\prime \sim D^{2/3}$.

Hence the leading boundary layer structure near the tip is~\cite{ref15}
\begin{equation}
v = g_- + D^{1/3} \hat{v} \left( \frac{r - r_{t}}{D^{2/3}}, 
\frac{\theta}{D^{1/3}} \right)
\end{equation}
Note that this is exactly the type of relationship that we need.  The 
tangential derivative
discontinuity of $v_{PS}$ can be explicitly canceled by the $\theta$ 
derivative of the second term;
if we iterate the equation perturbatively, we will find that the curvature 
of the PS solution needs to
be corrected by an amount of the form $D^{1/3} f (s / D^{1/3})$ for small 
arclengths and this
boundary layer will compensate for the jump in $\frac{\partial 
\kappa}{\partial s}$ across the tip. 
On the other hand, the normal derivative $\frac{\partial v}{\partial r} 
\mid_{r_{t}}$ is
actually of order $D^{-1/3}$ and hence appears to diverge in the 
Pelce-Sun solution.

To summarize, we have shown how to construct a singularity-free spiral 
field, starting from the
Pelce-Sun solution, by including the effects of small diffusion 
constant.  The fact that this
construction goes through without difficulty proves that the Pelce-Sun 
solution gives the correct
interface shape and concomitant rotation frequency in the $D \rightarrow 
0$ limit.  However, the
demonstration that the normal derivative of the field near the tip is 
actually divergent means that
the stability problem is subtle; since a perturbation to the interface 
will in general move the tip in
the radial direction, this induces large field changes which must be 
explicitly balanced by interfacial structure on the boundary layer
length scale $D^{1/3}$.  Work on this stability problem is in progress.

\acknowledgements
HL is supported in part by NSF Grant DMR94-15460. DAK is supported in
part by the Israel Science Foundation.


\begin{references}
\bibitem{ref1} For a review, see A. T. Winfree, {\em Chaos} {\bf 1}, 303 (1991).
\bibitem{ref2} D. Barkley, M. Kness and L. S. Tuckerman, {\em Phys. Rev.}
{\bf A42}, 2489 (1990); D. Barkley, {\em Phys. Rev. Lett} {\bf 68}, 2090 
(1994).
\bibitem{ref3} A. Karma, {\em Phys. Rev. Lett.} {\bf 65}, 2824 (1990).
\bibitem{ref4}Z. Nagy-Ungvarai, J. Ungvarai, and S. C. Muller, {\em Chaos}
{\bf 3}, 15 (1993).
\bibitem{ref5} G. Li, Q. Ouyang, V. Petrov and H. L. Swinney, {\em Phys.
Rev. Lett.} {\bf 77}, 2105 (1996).
\bibitem{ref6} ``Simulation of Spiral Interface in a Free-Boundary Limit", 
I. Mitkov,  I. Aranson and D. Kessler, {\em Phys. Rev E} to appear.
\bibitem{ref7} ``Spirals in Excitable Media II: Meandering Transition in 
the Free-Boundary Limit", D. Kessler and R.
Kupferman, {\em Physica D} to appear. 
\bibitem{ref7'} D. Kessler and H. Levine, {\em Physica} {\bf D49}, 90 (1991).
\bibitem{ref8} P. Pelce and J. Sun, {\em Physica} {\bf D48}, 353 (1991).
\bibitem{ref9}J. J. Tyson and J. P. Keener, {\em Physica} {\bf D32}, 327 (1988).
\bibitem{ref10}P. C. Fife in ``Non-equilibrium Dynamics in Chemical Systems",
C. Vidal and A. Pacault, eds. Springer (New York, 1984).
\bibitem{ref11}D. Kessler and R. Kupferman, {\em Physica} {\bf  D97}, 509 (1996).
\bibitem{ref10'}D. Kessler, H. Levine and W. N. Reynolds, {\em Physica}
{\bf D70}, 115 (1994).
\bibitem{ref12} There is one (unsuccessful) attempt to carry out such an 
analysis; see
P. Pelce and J. Sun, {\em Physica} {\bf D63}, 273 (1993). More recently,
(M. Falcke and H. Levine, in preparation), it has been shown that the
stability calculation can in fact be done if one makes certain assumptions
regarding the correct boundary conditions on the perturbed
interface near the tip. These assumptions could only be checked by a
regularized theory of the type we are constructing in this paper.
\bibitem{ref13} See e.g. ``Table of integrals, series, and products'', by 
I. S. Gradshteyn and I.  M. Ryzhik.  4th ed., Academic Press, 1965.
\bibitem{ref14} One can easily show that eq. (20) gives rise to a one-sided
exponential boundary layer by changing variables of integration. We should
note that for this boundary layer, the other term 
$\hat{n} ' \cdot \nabla ' G$
in eq. (8) is also relevant, since the derivative acting on the inner
variable (of order $D$) exactly compensates for the explicit extra factor of
$D$.
\bibitem{ref15} Again a quantitative treatment must include the
$\hat{n} ' \cdot \nabla ' G$, since there is no $D^{1/3}$ factor and the
normal derivative pulls down a factor of $D^{-2/3}$, which thus compensates
for the explicit extra factor of $D$ in eq. (8).
\end{references}
\end{document}